\documentclass[aps,prb,twocolumn,superscriptaddress,showpacs]{revtex4}
\usepackage{hyperref}
\usepackage[dvips]{graphicx}
\usepackage{amssymb,amsmath,amsfonts}
\usepackage[T1]{fontenc}

\usepackage{tikz}
\usetikzlibrary{positioning}

\newcommand*\semidirectsum{%
  \raisebox{-1pt}
{%
  \begin{tikzpicture}[scale=0.11]
\draw[line width=0.035 em, color=black] (0,0) -- (180:1);
\draw[line width=0.035 em, color=black] (0,0) -- (0:1);
\draw[line width=0.035 em, color=black] (0,0) -- (90:1);
\draw[line width=0.035 em, color=black] (40:1) arc (40:90:1);
\draw[line width=0.035 em, color=black] (40:1) arc (40:-90:1);
\draw[line width=0.035 em, color=black] (0,0) -- (-90:1);
\end{tikzpicture}}}

\begin{document}

% Use the \preprint command to place your local institutional report
% number in the upper righthand corner of the title page in preprint mode.
% Multiple \preprint commands are allowed.
% Use the 'preprintnumbers' class option to override journal defaults
% to display numbers if necessary
%\preprint{}

%Title of paper
\title{QED$_{2+1}$ in graphene: symmetries of Dirac equation in $2+1$\ dimensions}

%%%%%%%%%%%%%%%%%%%%%%%%%%%%%%%%%%%%%%%%%%%%%%%%%%%%%%%%%%%%%%%%%%%%%
%% The document title should be given as usual. Some journals require
%% a running title from the author: this should be supplied as an
%% optional argument to \title.
%%%%%%%%%%%%%%%%%%%%%%%%%%%%%%%%%%%%%%%%%%%%%%%%%%%%%%%%%%%%%%%%%%%%%

\author{P. Kosi\'nski}
\affiliation{Department of Theoretical Physics and Computer Science, University of Lodz, Pomorska 149/153, 90-236 Lodz, Poland}
\author{P. Ma\'slanka}
\affiliation{Department of Theoretical Physics and Computer Science, University of Lodz, Pomorska 149/153, 90-236 Lodz, Poland}
\author{J. S\l awi\'{n}ska}
\email[]{jagoda.slawinska@uni.lodz.pl}
\affiliation{Department of Theoretical Physics and Computer Science, University of Lodz, Pomorska 149/153, 90-236 Lodz, Poland}
\affiliation{Solid State Physics Department, University of Lodz, Pomorska 149/153, 90-236 Lodz, Poland}
\author{I. Zasada}
\affiliation{Solid State Physics Department, University of Lodz, Pomorska 149/153, 90-236 Lodz, Poland}

%\email[]{Your e-mail address}
%\homepage[]{Your web page}
%\thanks{}
%\altaffiliation{}

%Collaboration name if desired (requires use of superscriptaddress
%option in \documentclass). \noaffiliation is required (may also be
%used with the \author command).
%\collaboration can be followed by \email, \homepage, \thanks as well.
%\collaboration{}
%\noaffiliation
\vspace{.15in}

\begin{abstract}
It is well-known that the tight-binding Hamiltonian of graphene describes the low-energy excitations that appear to be massless chiral Dirac fermions. Thus, in the continuum limit one can analyze the crystal properties using the formalism of quantum electrodynamics in 2+1 dimensions (QED$_{2+1}$) which provides the opportunity to verify the high energy physics phenomena in the condensed matter system. We study the symmetry properties of 2+1-dimensional Dirac equation, both in the noninteracting case and in the case with constant uniform magnetic field included in the model. The maximal symmetry group of the massless Dirac equation is considered by putting it in the Jordan block form and determining the algebra of operators leaving invariant the subspace of solutions. It is shown that the resulting symmetry operators expressed in terms of Dirac matrices cannot be described exclusively in terms of $\gamma$ matrices (and their products) entering the corresponding Dirac equation. It is a consequence of the reducibility of the considered representation in contrast to the 3+1-dimensional case. Symmetry algebra is demonstrated to be a direct sum of two gl(2,C) algebras plus an eight-dimensional abelian ideal. Since the matrix structure which determines the rotational symmetry has all required properties of the spin algebra, the pseudospin related to the sublattices (M. Mecklenburg and B. C. Regan, Phys. Rev. Lett. 106, 116803 (2011)) gains the character of the real angular momentum, although the degrees of freedom connected with the electron's spin are not included in the model. This seems to be graphene's analogue of the phenomenon called "spin from isospin" in high energy physics.
\end{abstract}

\maketitle

\section{Introduction}
Graphene\cite{review} is a single layer of carbon atoms arranged into a planar honeycomb lattice. It has attracted a considerable attention since its fabrication in 2004\cite{novoselov_science}, due to its unique attributes being a consequence of charge conjugation symmetry between positive and negative charge carriers, which have the internal degree of freedom analogous to the chirality defined in the quantum field theory\cite{gusynin}. This similarity between condensed matter physics and the quantum electrodynamics (QED) makes graphene a test-bed for experimental verification of phenomena well-known in the particle physics\cite{katsnelson_bridge}. For example, the Klein paradox can be observed as the electron propagation through potential barriers with the graphene playing the role of an effective medium\cite{katsnelson_chiral}; the Zitterbewegung effect restricts the minimum conductivity to the order of conductance quantum $e^{2}/h$ (Ref.\onlinecite{novoselov_letter}), while the anomalous quantum Hall effect\cite{zhang_berry} can be related with the index theorem. 

Such exciting properties and perspectives are a direct consequence of the fact that the low-energy properties of electrons in graphene can be described by the model based on the continuum limit ($a\rightarrow0$) of the tight binding approximation\cite{semenoff, gusynin}, which obeys a relation formally identical to the 2+1 dimensional Dirac equation with the holes and the pseudospin states of the A and B sublattices being the counterparts of the positrons and the spin, respectively\cite{spin_lessons}. The remarkable insight into the origin of graphene's uniqueness can be achieved by the studies of the symmetries of the model. Recently, much attention has been paid to the close relation between spatial symmetries and the existence of the Dirac fermions which can be present even in the chemisorbed graphene with the defects distributed with specific symmetry\cite{prbrevised1}. This hypothesis have been also verified by means of density functional theory (DFT) calculations\cite{prbrevised2}. Moreover, the effects of symmetry breaking have been widely studied, for example, due to the mass term\cite{gusynin} or valley dependent vector potential\cite{beenakker}. The latter examples are extremely important for applications, since real graphene systems always interact with surrounding environment which disturbs its exceptional electronic properties\cite{review, hbn, holendrzy, mucha}.

In this context, it is useful to derive the fundamental symmetries of the model using the same mathematical procedures that are commonly applied to study a 3+1 Dirac equation\cite{nikitin}. The main difference between the two-dimensional system of graphene's fermions and the 3+1-dimensional system is that the Dirac equation of the latter is based on the four-dimensional irreducible representation of Clifford algebra. In contrast, in the analysis of the two-dimensional case one also uses the four-dimensional representation, but, as can be easily demonstrated, it becomes reducible. The Clifford algebra for three-dimensional space-time possesses two inequivalent irreducible representations which are both two-dimensional. Thus, the representation used to study the model of graphene is the direct sum of the irreducible ones. 

The purpose of this paper is to study the symmetries of the 2+1 massless Dirac equation and to clarify the consequences of the reducibility of the matrices representing the Clifford algebra. Understanding the symmetries of the underlying dynamics is especially profitable in the context of the possibilities of experimental verification of 2+1 QED-like phenomena in the condensed matter physics as well as tuning of the graphene's properties for applications. Due to the significant role of the experiments concerning the anomalous quantum Hall effect in graphene, it is also useful to study the case of 2+1 system in the presence of magnetic field. Moreover, it is the only type of interaction that does not disturb the algebraic structure of symmetries of this system and does not lead to any qualitative changes in the spectrum (in the sense that it keeps the energy gap between particles and holes intact).

In this paper we have determined the algebra of the corresponding symmetry operators using the procedures elaborated in the case of four-dimensional Dirac equation in Ref.\onlinecite{nikitin}, both for free case as well as the case with the constant uniform magnetic field. The convenient way to treat the latter case is to make use of the supersymmetry inherent to the problem.

The paper is organized as follows: in Sec.II  the "on-shell" symmetry generators for free massless Dirac equation are introduced and the explicit form of its Lie algebra generators is found. In Sec.III  the extended model with the magnetic field included is considered and compared to the noninteracting case described in Sec. II. The conclusions and perspectives are discussed in Sec. IV.

\section{The symmetries of the free massless Dirac equation}
We consider the free massless Dirac equation in momentum representation:
 \begin{eqnarray}
\gamma ^\mu p_\mu \Psi =0 \label{w1}
\end{eqnarray}
where $\gamma$ matrices are given in the following representation:
\begin{eqnarray}
\gamma ^0=\left[\begin{array}{cc}
0 & I \\
I & 0\end{array}\right], \;\;\;
\gamma ^i=\left[\begin{array}{cc}
0 &\sigma ^i \\
-\sigma ^i & 0\end{array}\right],\;\;\;\gamma ^5=i\gamma^0 \gamma^1 \gamma ^2\gamma^3 \label{w2}
\end{eqnarray}
with $i=1,2,3$ (although only $\gamma^0 ,\gamma^1 ,\gamma ^2$\ enter the considered representation of the Clifford algebra). Eq.(\ref{w1}) 
implies $p_0=\pm \mid \vec{p} \mid $; thus we can write 
 \begin{eqnarray}
\gamma ^\mu p_\mu  =\pm \mid \vec{p} \mid (\gamma ^0\pm n_i \gamma^i )\label{w3}
\end{eqnarray}
where $n_i=p_i/\mid \vec{p} \mid$ with $i=1,2$, is the unit vector in the direction of momentum. We can restrict our analysis to the case with
$p_0=\mid \vec{p} \mid$. The minus sign in eq. (\ref{w3}) can be accounted for by reversing the sign of the momentum. Imposing the condition 
$\mid \vec{p} \mid\not=0$\ one is left with the operator:
 \begin{eqnarray}
L\equiv \gamma ^0+n_i\gamma ^i=\left[\begin{array}{cc}
0 &I-n_i\sigma ^i \\
I+n_i\sigma ^i & 0\end{array}\right] \label{w4}
\end{eqnarray}
Next, we can represent $\Psi$ by the Weyl spinors:
 \begin{eqnarray}
\Psi =\left[\begin{array}{cc}
\Psi _+ \\
 \Psi _-\end{array}\right], \;\;\; \Psi _\pm =\frac{1\pm\gamma ^5}{2}\Psi  \label{w5}
\end{eqnarray}
Then eq.(\ref{w1}) reduces to:
 \begin{eqnarray}
\Psi _\pm =\mp n_i\sigma ^i\Psi _\pm  \label{w6}
\end{eqnarray}
For any spinor $\Psi $\ we put $(\epsilon ,\epsilon '=\pm $)
 \begin{eqnarray}
\Psi ^{\epsilon '}_\epsilon =\left(\frac{1+\epsilon \gamma ^5}{2}\right)
\left(\frac{1+\epsilon 'n_i\sigma ^i}{2}\otimes I\right)\Psi  \label{w7}
\end{eqnarray}
and $\Psi ^{\epsilon '}_\epsilon $\ form a new basis given by the following expressions:
 \begin{eqnarray}
e_1=\frac{1}{2} \Psi ^-_-,\quad\; e_2=\Psi ^-_+\\
e_3=\frac{1}{2}\Psi ^+_+,\quad\; e_4=\Psi ^+_-
\end{eqnarray}
In this basis the operator $L$\ acquires the Jordan block form:
 \begin{eqnarray}
\tilde{L}=\left[\begin{array}{cccc}
0 & 0 & 0 & 0 \\
1 & 0 & 0 & 0 \\
0 & 0 & 0 & 0 \\
0 & 0 & 1 & 0 \end{array}\right] \label{w8}
\end{eqnarray}
We define the symmetry operators as the matrices $S$\ preserving the null eigenspace of $L$:
\begin{eqnarray}
\tilde{L}\Psi =0 \;\;\Longrightarrow \;\;\tilde{L} S\Psi =0 \label{w9} 
\end{eqnarray}
It is straightforward to compute the general form of $S$:
\begin{eqnarray}
S=\left[\begin{array}{cccc}
a_{11} &0 & a_{12}& 0 \\
x_{11} & b_{11} & x_{12} & b_{12} \\
a_{21} & 0 & a_{22} & 0 \\
x_{21}& b_{21} & x_{22} & b_{22} \end{array}\right]; \label{w10}
\end{eqnarray}
We can analyze the properties of S in terms of its Lie algebra's structure. First, it is worthwhile to note that there are two commuting $gL(2,\bold C)$\ subalgebras consisting of the matrices given in the following form:
 \begin{eqnarray}
\left[\begin{array}{cccc}
a_{11} &0 & a_{12}& 0 \\
0 & 0 & 0 & 0 \\
a_{21} & 0 &a_{22} & 0 \\
0 & 0 & 0 & 0 \end{array}\right] \;\;
\mathrm{and} \;\;\;
 \left[\begin{array}{cccc}
0 & 0 & 0 & 0 \\
0 & b_{11} & 0 & b_{12} \\
0 & 0 & 0 & 0 \\
0& b_{21} & 0 & b_{22} \end{array}\right] \label{w11}
\end{eqnarray}
Moreover, there is a four-dimensional complex abelian algebra $A(4,\bold C)$:
 \begin{eqnarray}
S=\left[\begin{array}{cccc}
0 & 0 & 0 & 0 \\
x_{11} & 0 & x_{12} & 0 \\
0 & 0 & 0 & 0 \\
x_{21} & 0 & x_{22} & 0 \end{array}\right] \label{w12}
\end{eqnarray}
Under the adjoint action $A(4,\bold C)$\ provides the representation of both $gL(2,\bold C)$\ subalgebras. One can conclude that the symmetry algebra has the following structure:
 \begin{eqnarray}
S=(gL(2,\bold C) \oplus gL(2,\bold C))\semidirectsum\,  A(4,\bold C) \label{w13}
\end{eqnarray}
where  \semidirectsum\,  denotes the semi-direct sum. Explicitely, choosing the standard basis one finds:
\begin{eqnarray}
&&[\tilde{A_{ij}},\tilde{A_{mn}}]=\delta _{jm}\tilde{A_{in}}- \delta _{in}\tilde{A_{mj}} \nonumber \\
&&[\tilde{B_{ij}},\tilde{B_{mn}}]=\delta _{jm}\tilde{B_{in}}- \delta _{in}\tilde{B_{mj}} \nonumber \\
&&[\tilde{X_{ij}},\tilde{X_{mn}}]=0  \nonumber \\
&&[\tilde{A_{ij}},\tilde{X_{mn}}]=-\delta _{in}\tilde{X_{mj}} \nonumber \\
&&[\tilde{B_{ij}},\tilde{X_{mn}}]=\delta_ {jm}\tilde{X_{in}}  \label{w14}
\end{eqnarray}
One can return to the natural basis (\ref{w4}) using the similarity transformation $L=V\tilde{L}V^{-1}$\ with the matrix:
 \begin{eqnarray}
V=\left[\begin{array}{cccc}
0 & n_- & 0 & 0 \\
0 & -1 & 1 & 0 \\
0 & 0 & 0 & n_- \\
-1 & 0 & 0 & 1\end{array}\right] \label{w115}
\end{eqnarray} 
where $n_\pm =n_1\pm in_2$.\\
Finally, simple calculations lead to the following result for generators of symmetry algebra:
 \begin{eqnarray}
&&A_{11}=\frac{1}{4}(1-\gamma ^5)(1+\gamma ^0\gamma ^3) - \frac{1}{4}n_+\gamma ^0(1+\gamma ^5)\gamma ^0\gamma ^-\nonumber \\
&&A_{12}=-\frac{1}{4}\gamma ^0(1+\gamma ^5)(1-\gamma ^0\gamma ^3) + \frac{1}{4}n_+\gamma ^0(1+\gamma ^5)\gamma ^0\gamma ^-\nonumber \\ 
&&A_{21}=-\frac{1}{4}\gamma ^0(1-\gamma ^5)(1+\gamma ^0\gamma ^3) - \frac{1}{4}n_+\gamma ^0(1-\gamma ^5)\gamma ^0\gamma ^-\nonumber \\
&&B_{11}=\frac{1}{4}(1+\gamma ^5)(1+\gamma ^0\gamma ^3) - \frac{1}{4}n_+(1+\gamma ^5)\gamma ^0\gamma ^-\nonumber \\
&&B_{12}=\frac{1}{4}\gamma ^0(1-\gamma ^5)(1-\gamma ^0\gamma ^3) + \frac{1}{4}n_+\gamma ^0(1-\gamma ^5)\gamma ^0\gamma ^-\nonumber \\ 
&&B_{21}=\frac{1}{4}\gamma ^0(1+\gamma ^5)(1+\gamma ^0\gamma ^3) - \frac{1}{4}n_+\gamma ^0(1+\gamma ^5)\gamma ^0\gamma ^-\nonumber \\
&&B_{22}=\frac{1}{4}(1-\gamma ^5)(1-\gamma ^0\gamma ^3) + \frac{1}{4}n_+\gamma ^0(1-\gamma ^5)\gamma ^0\gamma ^-\nonumber \\
&&X_{11}=\frac{1}{4}\gamma ^0(1-\gamma ^5)(1-\gamma ^0\gamma ^3) + \frac{1}{4}n_+\gamma ^0(1-\gamma ^5)\gamma ^0\gamma ^-\nonumber \\
&&\;\;\;\;\;\;\;+\frac{1}{4}\gamma ^0(1-\gamma ^5)(1+\gamma ^0\gamma ^3) + \frac{1}{4}n_-\gamma ^0(1-\gamma ^5)\gamma ^0\gamma ^+\nonumber \\
&&X_{12}=\frac{1}{4}\gamma ^0(1+\gamma ^5)(1+\gamma ^0\gamma ^3) + \frac{1}{4}n_-(1+\gamma ^5)\gamma ^0\gamma ^+\nonumber \\
&&\;\;\;\;\;\;\;-\frac{1}{4}(1+\gamma ^5)(1-\gamma ^0\gamma ^3) - \frac{1}{4}n_+(1+\gamma ^5)\gamma ^0\gamma ^-\nonumber \\
&&X_{21}=\frac{1}{4}(1-\gamma ^5)(1-\gamma ^0\gamma ^3) + \frac{1}{4}n_-(1-\gamma ^5)\gamma ^0\gamma ^+\nonumber \\
&&\;\;\;\;\;\;\;+\frac{1}{4}n_+\gamma ^0(1+\gamma ^5)\gamma ^0\gamma ^- + \frac{1}{4}n_-(1-\gamma ^5)(1+\gamma ^0\gamma ^3)\nonumber \\
&&X_{22}=\frac{1}{4}\gamma ^0(1+\gamma ^5)(1+\gamma ^0\gamma ^3) + \frac{1}{4}n_-\gamma ^0(1-\gamma ^5)\gamma ^0\gamma ^+\nonumber \\
&&\;\;\;\;\;\;\;-\frac{1}{4}n_+\gamma ^0(1+\gamma ^5)\gamma ^0\gamma ^- + \frac{1}{4}n_-\gamma ^0(1+\gamma ^5)(1-\gamma ^0\gamma ^3)\nonumber\\
&&\;\;\;\;\;\;\;\;\;\;\;\;\;\;\;  \label{w116}
\end{eqnarray}
where we denoted $\gamma ^\pm =\gamma ^1\pm i\gamma ^2$.\\
The explicit form of the generators of the symmetry algebra in the representation given by eq. (\ref{w2}) reads:
\begin{eqnarray}
&&A_{11}=\left[\begin{array}{cccc}
0 & 0 & 0 & 0 \\
0 & 0 & 0 & 0 \\
0 & 0 & 0 & 0 \\
0 & 0 & -n_+ & 1 \end{array}\right]  \;\;\;\;\;\;\;\;\;\;\;\;
A_{12}=\left[\begin{array}{cccc}
0 & 0 & 0 & 0 \\
0 & 0 & 0 & 0 \\
0 & 0 & 0 & 0 \\
-n_+ & -1 & 0 & 0 \end{array}\right] \nonumber\\
&&A_{21}=\left[\begin{array}{cccc}
0 & 0 & 0 & 0 \\
0 & 0 & n_+ & -1 \\
0 & 0 & 0 & 0 \\
0 & 0 & 0 & 0 \end{array}\right] \;\;\;\;\;\;\;\;\;\;\;\;
A_{22}=\left[\begin{array}{cccc}
0 & 0 & 0 & 0 \\
n_+ & 1 & 0 & 0 \\
0 & 0 & 0 & 0 \\
0 & 0 & 0 & 0 \end{array}\right] \nonumber\\
&&B_{11}=\left[\begin{array}{cccc}
1 & 0 & 0 & 0 \\
-n_+ & 0 & 0 & 0 \\
0 & 0 & 0 & 0 \\
0 & 0 & 0 & 0\end{array}\right] \;\;\;\;\;\;\;\;\;\;\;\;
B_{12}=\left[\begin{array}{cccc}
0 & 0 & 1 & 0 \\
0 & 0 & -n_+ & 0 \\
0 & 0 & 0 & 0 \\
0 & 0 & 0 & 0 \end{array}\right] \nonumber\\
&&B_{21}=\left[\begin{array}{cccc}
0 & 0 & 0 & 0 \\
0 & 0 & 0 & 0 \\
1 & 0 & 0 & 0 \\
n_+ & 0 & 0 & 0 \end{array}\right] \;\;\;\;\;\;\;\;\;\;\;\;\;\;\;
B_{22}=\left[\begin{array}{cccc}
0 & 0 & 0 & 0 \\
0 & 0 & 0 & 0 \\
0 & 0 & 1 & 0 \\
0 & 0 & n_+ & 0 \end{array}\right] \nonumber\\
&&X_{11}=\left[\begin{array}{cccc}
0 & 0 & 1 & -n_- \\
0 & 0 & -n_+ & 1 \\
0 & 0 & 0 & 0 \\
0 & 0 & 0 & 0 \end{array}\right] \;\;\;\;\;\;\;
X_{12}=\left[\begin{array}{cccc}
1 & n_- & 0 & 0 \\
-n_+ & -1 & 0 & 0 \\
0 & 0 & 0 & 0 \\
0 & 0 & 0 & 0 \end{array}\right] \nonumber\\
&&X_{21}=\left[\begin{array}{cccc}
0 & 0 & 0 & 0 \\
0 & 0 & 0 & 0 \\
0 & 0 & 1 & -n_- \\
0 & 0 & n_+ & -1 \end{array}\right] \;\;\;\;\;\;\;\;\;\;
X_{22}=\left[\begin{array}{cccc}
0 & 0 & 0 & 0 \\
0 & 0 & 0 & 0 \\
1 & n_- & 0 & 0 \\
n_+ & 1 & 0 & 0 \end{array}\right] \nonumber\\
&&\;\;\;\;\;\;\;\;\;\;\;\;\;\;\;\;  \label{w117}
\end{eqnarray}
\normalsize

Let us note that, due to the fact that the Dirac matrices $\gamma_{0}, \gamma_{1}, \gamma_{2}$ generate a reducible algebra, one cannot express all symmetry operators in terms of them only\cite{diracmatrices}; instead, as can be seen from eq. (\ref{w116}), one has to add $\gamma_{3}$ matrix to obtain irreducible algebra containing all necessary matrices.

\section{Dirac equation in a constant uniform magnetic field}
We consider the 2+1-dimensional Dirac equation in the constant uniform magnetic field $B$. The general symmetry pattern in this case is the result of the following properties: (i) the 2+1-dimensional Dirac equation based on the irreducible representation of Clifford algebra exhibits supersymmetry\cite{thaller}, (ii) the representation of the Clifford algebra under consideration is reducible and is the direct sum of both inequivalent irreps.

We include the electromagnetic field through a minimal coupling. A convenient gauge choice, which assures a constant uniform magnetic field, is $A_\mu =(0,-Bx^2/2,Bx^1/2)$. The Dirac equation takes now the form:
\begin{eqnarray}
\left[i\gamma ^0(\partial _0-i\mu )+iv_F\gamma ^1D_1+iv_F\gamma ^2D_2\right]\Psi =0 \label{w18} 
\end{eqnarray}
where we define:
\begin{eqnarray}
D_1=\partial _1-\frac{ieBx^2}{2},\;\;\;\;\;\;D_2=\partial _2+\frac{ieBx^1}{2} \label{w19} 
\end{eqnarray}
Inserting $\Psi = exp(-iEx^0)\Phi$\ to (\ref{w18}) leads to the following Hamiltonian form of the Dirac equation:
\begin{eqnarray}
(E+\mu )\Phi (x)=-iv_F\left(\gamma ^0\gamma ^1D_1+\gamma ^0\gamma ^2D_2\right)\Phi (x) \label{w20} 
\end{eqnarray}
In terms of the two-component spinors $\Phi  =\left[\begin{array}{c}
\chi  \\
\kappa \end{array}\right]$\ the equation (\ref{w20}) reads
\begin{eqnarray}
&&(E+\mu )\chi =-iv_F\left(\sigma ^kD_k\right)\chi \nonumber\\
&&(E+\mu )\kappa  =iv_F\left(\sigma ^kD_k\right)\kappa   \label{w21} 
\end{eqnarray}
This is the set of decoupled equations for upper and lower components. Thus, the Dirac Hamiltonian has the form:
 \begin{eqnarray}
H=\left[\begin{array}{cc}
H_+ & 0 \\
0 & H_-\end{array}\right]  \label{w22}
\end{eqnarray}
with
\vskip 0.3 cm
\footnotesize
 \begin{eqnarray}
H_{\pm}=\mp iv_F\left[\begin{array}{cc}
0 & \partial _1-i\frac{eB}{2}x^2-i\partial _2+\frac{eB}{2}x^1 \\
\partial _1-i\frac{eB}{2}x^2+i\partial _2-\frac{eB}{2}x^1 & 0\end{array}\right]\nonumber\end{eqnarray}
\normalsize
\begin{eqnarray}
\;\;\;\label{w23}
\end{eqnarray}
Below we will follow the discussion given in Ref.\onlinecite{haldane}. It is useful to introduce the complex variable:
$z=x^2+ix^1$\ and to define the operators:
\begin{eqnarray}
&&D=2\frac{\partial }{\partial \bar{z}}+\frac{eB}{2}z  \nonumber\\
&&D^*=-2\frac{\partial }{\partial z}+\frac{eB}{2}\bar{z}     \label{w24} 
\end{eqnarray}
It can be easily found that they obey the algebra: 
\begin{eqnarray}
[D,D^*]=2eB \label{w25} 
\end{eqnarray}
The angular momentum operator expressed in terms of the above defined complex variable reads:
\begin{eqnarray}
J=\bar{z}\frac{\partial }{\partial \bar{z} }-z\frac{\partial }{\partial z}+\frac{1}{2}\sigma ^3  \label{w26} 
\end{eqnarray}
The $\sigma^{3}$ - contribution is the so-called lattice spin and not the real electron spin (cf. Ref.\onlinecite{spin_lessons} and the discussion in Sec. IV below.)

The Hamiltonian $H_\pm $\ acquire a very simple form:
 \begin{eqnarray}
H_\pm =\mp v_F\left[\begin{array}{cc}
0 & D \\
D^* & 0\end{array}\right];  \label{w27}
\end{eqnarray}
This shows that the eigenvalue problem exhibits supersymmetry\cite{thaller}. Therefore, proceeding in the standard way we define the following vectors:
 \begin{eqnarray}
\Psi _n=(D^*)^n\Psi _0  \label{w28}
\end{eqnarray}
where $\Psi _0$\ obeys:
 \begin{eqnarray}
D\Psi _0=0 \label{w29}
\end{eqnarray}
It is now straightforward to check that $\Psi _n$\ obey the eigenvalue equations:
\begin{eqnarray}
&&D^*D\Psi _n=2neB\Psi _n \nonumber \\
&&DD^*\Psi _n= 2(n+1) eB\Psi _n  \label{w30}
\end{eqnarray}
It can be easily found that the eigenvalues and the eigenvectors of $H_+$\ have the following form:\small
 \begin{eqnarray}
&&H_+\chi _{n+}=v_F\sqrt{2neB}\chi _{n+},\;\;\;\chi _{n+}  =\left[\begin{array}{c}
-\sqrt{2neB}\Psi _{n-1} \\
\\
\Psi _n \end{array}\right], \nonumber \\
&&H_+\chi _{n-}=-v_F\sqrt{2(n+1)eB}\chi _{n-},\;\;\;\chi _{n-}  =\left[\begin{array}{c}
\Psi _n \\
\\
\frac{\Psi _{n+1}}{\sqrt{2(n+1)eB}} \end{array}\right]\nonumber\\
&&\label{w31}
\end{eqnarray}\normalsize
where $n=0,1,2,...$.
It should be noted that the spectrum of $H_+$\ is infinitely degenerate. Namely, $H_+$\ commutes with angular momentum $J$,
so a given eigenspace of $H_+$\ is spanned by the eigenvectors of $J$\ corresponding to arbitrary allowed values of angular momentum.
To explain this, one can observe that the equation (\ref{w29}) defines $\Psi _0$\ up to the arbitrary multiplicative factor depending on $z$. The general solution of the equation defining $\Psi _0$\ can be written as a linear combination of following eigenvectors:
\begin{eqnarray}
\Psi _0^m=z^me^{-\frac{eB}{4}z\bar{z}}  \label{w32}
\end{eqnarray}
They carry the orbital angular momentum $-m$\ and allows us to define, with the help of eq. (\ref{w28}) the towers of eigenvectors of $DD^*$\ and $D^*D$:
\begin{eqnarray}
\Psi_n^m=(D^*)^n\Psi _0^m \label{w33}
\end{eqnarray}
$\Psi _n^m$\ are eigenvectors of orbital angular momentum operator corresponding to the eigenvalue $n-m$. Accordingly, the complete set of eigenvectors of $H_+$ reads:
\begin{eqnarray}
\chi _{n+}^m  =\left[\begin{array}{c}
-\sqrt{2neB}\Psi _{n-1}^m \\
\\
\Psi _n^m \end{array}\right],\;\;\; \chi _{n-}^m  =\left[\begin{array}{c}
\Psi _n^m \\
\\
\frac{\Psi _{n+1}^m}{\sqrt{2(n+1)eB}} \end{array}\right], \label{w34}
\end{eqnarray}
The total angular momentum carried by $\chi _{n\pm }^m$\ equals $n-m\mp 1/2$.
For $H_-$\ one similarly obtain the following eigenvalues and eigenvectors:
 \begin{eqnarray}
&&H_-\kappa  _{n-}^m=-v_F\sqrt{2neB}\kappa  _{n-}^m,\;\;\;\kappa  _{n-}^m  =\left[\begin{array}{c}
-\sqrt{2neB}\Psi _{n-1}^m \\
\\
\Psi _n^m \end{array}\right]\nonumber \\
&&H_-\kappa _{n+}^m=v_F\sqrt{2(n+1)eB}\kappa  _{n+}^m,\;\;\;\kappa  _{n+}^m  =\left[\begin{array}{c}
\Psi _n^m \\
\\
\frac{\Psi _{n+1}^m}{\sqrt{2(n+1)eB}} \end{array}\right]\nonumber\\
&& \label{w35}
\end{eqnarray}
The total angular momentum carried by $\kappa _{n\pm }^m$\ equals $n-m\mp 1/2$. \\

Let us consider the subspace corresponding to definite energy, $E_n=v_F\sqrt{2neB}-\mu $\ and angular momentum $j-n-m-1/2$.
Dirac equation takes the form:
\begin{eqnarray}
L\left[\begin{array}{c}
X  \\
Y \end{array}\right] \equiv\left[\begin{array}{cc}
0 & E_n+\mu -H_- \\
E_n+\mu -H_+ & 0\end{array}\right]\left[\begin{array}{c}
X  \\
Y \end{array}\right]=0 \nonumber\\
&&\label{w36}
\end{eqnarray}
Two eigenvectors corresponding to the above energy and angular momentum have the following form:
\begin{eqnarray}
e_2=\left[\begin{array}{c}
-\sqrt{2neB}\Psi _{n-1}^m \\
\\
\Psi _n^m\\
\\
0\\
\\
0 \end{array}\right] \;\;\;\;\;e_4=\left[\begin{array}{c}
0\\
\\
0\\
\\
\Psi _{n-1}^m \\
\\
\frac{\Psi _n^m}{\sqrt{2neB}} \end{array}\right] \label{w37}
\end{eqnarray}
We define two further vectors forming together with $e_2$ and $e_4$\ a basis:
\begin{eqnarray}
e_1=\left[\begin{array}{c}
0 \\
\\
0\\
-\Psi _{n-1}^m\\
\\
\frac{\Psi _n^m}{2\sqrt{2neB}}\end{array}\right] \;\;\;\;\;e_3=\left[\begin{array}{c}
\frac{\Psi _{n-1}^m}{2\sqrt{2neB}}\\
\\
\frac{\Psi _n^m}{4neB}\\
\\
0 \\
\\
0 \end{array}\right] \label{w38}
\end{eqnarray}
One can observe that $e_1,e_2,e_3,e_4$\ span an invariant subspace for the operator $L$\ in which $L$\ takes the form given by (\ref{w8}).
Therefore the symmetry algebra of Dirac equation restricted to the subspace of definite energy and angular momentum is:
 \begin{eqnarray}
S=(gL(2,\bold C) \oplus gL(2,\bold C))\semidirectsum\, A(4,\bold C) \label{w39}
\end{eqnarray}

We have found that the symmetry algebra coincides with the one corresponding to the free case. However, both cases differ by the choice of commuting variables defining the relevant subspace. In the free case these were the momentum components while here we consider the subspaces of given energy and angular momentum. 

\section{Discussion and perspectives}
We have analyzed the "on-shell" symmetries of 2+1 dimensional Dirac equation for the free particle as well as the one interacting with uniform magnetic field. In both cases the symmetry algebra appears to be the same (cf. eq. (\ref{w39})). However, it refers to the subspaces defined by different choices of commuting operators (the momentum components in the free case and energy and angular momentum in the case of uniform magnetic field).

The main difference between the standard 3+1-dimensional Dirac equation and the one considered here is that in the former the $\gamma$ matrices form the (unique up to equivalence) irreducible representation of Clifford algebra. On the contrary, in the 2+1-dimensional case there are two, both two-dimensional, inequivalent  irreducible representations. When writing out the Dirac equation we use both of them forming the direct sum which provides four-dimensional reducible representation. As a result, not all symmetry operators can be constructed out of $\gamma$ matrices (and their products) entering the relevant Dirac equation; one has to use, additionally, the $\gamma_{3}$ matrix which is absent from the relevant Dirac equation.

Let us note that the matrix structure which determines the rotational symmetry has all properties of spin algebra, in spite of the fact that the internal degrees of freedom come from the existence of two sublattices rather than from electron spin degrees of freedom which are neglected in the considered approximation. In fact, it has been argued\cite{spin_lessons} that the pseudospin, arising from the degeneracy introduced by the honeycomb lattice's two atomic sites per unit cell, has properties of real angular momentum. The form of the solutions of Dirac equation in magnetic field which we have used to find the symmetry algebra supports this point of view: we are dealing with the solution describing spin-$\frac{1}{2}$ particles in magnetic field. The problem of "lattice spin" seems to be worth further considerations. In high energy physics the emergence of spin from other degrees of freedom is known phenomenon called "spin from isospin"\cite{hep1,hep2}. For example, one can consider isospin gauge theory with bosonic matter of isospin $\frac{1}{2}$ (as well as neutral Higgs boson). It possesses the monopole solution. Quantizing the theory in one-monopole sector one finds that, due to the very presence of the monopole the SU(2)$_{\mathrm{rotations}}\times$SU(2) isospin symmetry is broken down to its diagonal subgroup which is the symmetry of monopole configuration. The relevant generators are the sums of ordinary rotations and isorotations. As a result, with bosons of half-integer isospin one obtains half-integer spin, in spite of the fact, that there are no elementary fermions involved. It would be interesting to interpret the results of Ref.\onlinecite{spin_lessons} in a similar spirit.

\begin{acknowledgments}
This work is financially supported by Polish Ministry of Science and Higher Education in the frame of Grant No. N~N202~204737 (P.K) and No. N~N202~086040 (J.S).
\end{acknowledgments}
\end{document}